\documentclass[journal]{IEEEtran}
\usepackage{cite}
\usepackage{amsmath,amssymb,amsfonts}
\usepackage{algorithmic}
\usepackage{graphicx}
\usepackage{textcomp}
\usepackage{xcolor}
\usepackage{multirow}
\usepackage{multicol}
\def\BibTeX{{\rm B\kern-.05em{\sc i\kern-.025em b}\kern-.08em
    T\kern-.1667em\lower.7ex\hbox{E}\kern-.125emX}}

\usepackage{geometry}
\makeatletter

\def\ps@IEEEtitlepagestyle{%
	\def\@oddfoot{\mycopyrightnotice}%
	\def\@evenfoot{}%
}
\def\mycopyrightnotice{%
	{\footnotesize  978-1-4799-6773-5/14/\$31.00 \textcopyright2017 Crown\hfill}
	\gdef\mycopyrightnotice{}
}



\makeatletter
\newcommand*\titleheader[1]{\gdef\@titleheader{#1}}
\AtBeginDocument{%
	\let\st@red@title\@title%
	\def\@title{%
		\bgroup\normalfont\large\centering\@titleheader\par\egroup
		\vskip0.1em\st@red@title}
}
\makeatother

\title{On Session Continuation among Slices for Inter-Slice Mobility 
	Support in 3GPP Service-based Architecture}
\titleheader{\large DOI: 10.1109/PIMRC48278.2020.9217332 \\
	\scriptsize \copyright 2020 IEEE. Personal use of this material is 
	permitted. 
	Permission from IEEE 
	must be obtained for all other uses, in any current or future media, 
	including reprinting/republishing this material for advertising or 
	promotional purposes, creating new collective works, for resale or 
	redistribution to servers or lists, or reuse of any copyrighted component 
	of this work in other works.}

\begin{document}
\author{Muhammad~Mohtasim~Sajjad\textsuperscript{*}, 
Dhammika~Jayalath\textsuperscript{*}, 
Yu-Chu~Tian\textsuperscript{\dag}, 
	and~Carlos~J.~Bernardos\textsuperscript{\textsection}\\
	\textsuperscript{*}School of Electrical Engineering and Robotics,\\
	\textsuperscript{\dag}School of Computer Science,\\
	\textsuperscript{*}\textsuperscript{\dag}Queensland University of 
	Technology, Brisbane, Australia\\
	\textsuperscript{\textsection}Universidad Carlos III de Madrid, Spain} 
%
%
\maketitle
\thispagestyle{empty}
\pagestyle{empty}
\begin{abstract}
	The 3GPP has provided its first standard specifications for network slicing 
	in the recent Release 15. The fundamental
	principles are specified which constitute the standard
	network slicing framework. These specifications, however, lack
	the session continuation mechanisms among slices, which is a
	fundamental requirement to achieve inter-slice mobility. In this
	paper, we propose three solutions which enable session continuation
	among slices in the current 3GPP network slicing framework.
	These solutions are based on existing, well-established
	standard mechanisms. The first solution is based on the Return
	Routability/Binding Update (RR/BU) procedure  of the popular
	Internet standard, Mobile IPv6 (MIPv6). 
	The second solution is based on the 3GPP standard GPRS Tunnelling Protocol 
	User Plane 
	(GTPv1-U), which establishes a GTP tunnel between previous and new slice 
	for session continuation. 
	The third solution is a hybrid
	solution of both
	MIPv6-RR/BU and GTPv1-U protocols. 
	We compare the performance
	of all these solutions through analytical modelling. 
Results show 
	that the GTPv1-U based and the hybrid MIPv6/GTPv1-U solutions promise lower 
	service 
	disruption 
	latency, however, incur higher resource utilization overhead compared to 
	MIPv6-RR/BU and 3GPP standard PDU Session 
	Establishment process. 

\end{abstract}


\section{Introduction}
Network Slicing has emerged as a key 
enabling technology for 5G, which provides different 
services types over a common network infrastructure. 
The 3GPP has provided a baseline framework for network 
slicing in its recent Release 15, which is based on a 
novel Service-based Architecture (SBA).
%
%
The ongoing 3GPP standardization on 5G provides 
more clarity on 
real-world semantics, 
requirements and challenges of a sliced mobile network. Among these, the 
challenge of managing mobility among different SBA slices, or Inter-Slice 
Handovers 
(ISHOs) is 
an important problem. 

As stated in \cite{Foukas2017, Li2017}, 
inter-slice mobility management 
requires new solutions. Most of the existing literature on mobility management 
in a sliced network 
focuses on managing slice-aware horizontal \cite{Zhang2017, Wen2018} or 
vertical handovers \cite{Manieshwar2019}. The problem of inter-slice mobility 
management is considered only by a limited studies. 
An ISHO solution in 
\cite{Yousaf2017} proposes virtualized mobility management applications, 
which can manage ISHOs in a softwarized network. Some other 
ISHO 
solutions consider specific vehicular communications use case of network 
slicing \cite{Mouawad2019, Kammoun2020}. 
However, none of the existing works have considered the 
network slicing framework of the 3GPP SBA. As such, the problem of managing 
mobility between slices in 3GPP SBA remains an open issue. 

Inter-slice mobility management requires smooth session continuation or session 
transfer among slices when a user moves from one slice to another. Different 
session types including IP, Ethernet, and Unstructured are defined by 3GPP to 
support different service types. However, the session continuation support 
is available only between 3GPP and non-3GPP access networks and the 
LTE's Evolved Packet System (EPS) to 5G core (5GC). The current 3GPP 
specifications do not support session continuation among slices. 



Thus, in this 
paper, we focus on the problem of session continuation among slices. The 
sessions of type IP, and among these IPv6, are considered which are expected to 
support 
a wide spectrum of 5G use cases. We 
present three solutions which are based on existing, well-established standard 
mechanisms. The first solution is based on Return Routability/Binding Update 
(RR/BU)
procedure of the popular IETF (Internet Engineering Task Force) Mobile IPv6 
(MIPv6) protocol \cite{mipv6}. This solution, termed as MIPv6-RR/BU, is based 
on 
a proposed concept of 
\textit{Home 
Slice}, through which the User Equipment (UE) communicates with the external 
Data Network (DN) 
using MIPv6 signalling. Through MIPv6 signalling, the UE requests the DN to 
redirect the ongoing traffic to its
new 
slice. The second solution is based on the 3GPP's standard  user plane 
tunnelling 
protocol, called GPRS Tunnelling Protocol User Plane (GTPv1-U) 
\cite{gtpv1u}. 
In this 
solution, a new functional entity at data plane, named \textit{Inter-Slice 
Gateway 
(ISGW)}, 
is proposed 
which 
manages an inter-slice GTP tunnel. The third 
solution is a hybrid solution of MIPv6-RR/BU and GTPv1-U protocols, and is 
termed as MIPv6/GTPv1-U. In this 
solution, 
traffic tunnelling through GTP occurs temporarily until the MIPv6-RR/BU 
operations 
complete. Through 
analytical modelling, we have compared the inter-slice handover performance and 
have highlighted the pros and cons of each solution. 

The remainder of this paper is organized as follows. In Section II, we 
briefly review 
the network 
slicing framework for SBA as 
specified by the 
3GPP. In particular, the standard principles related to a UE's 
movement from one slice to another, are discussed. Thereafter, in Section III, 
we present the proposed solutions. Section IV, 
presents analytical models, and the comparative analysis based on these models 
is given in Section V. Finally, we conclude this paper in Section 
VI. 

\section{3GPP Network Slicing Framework}
\begin{figure}[tbp]
	\centerline{\includegraphics[scale=0.1]{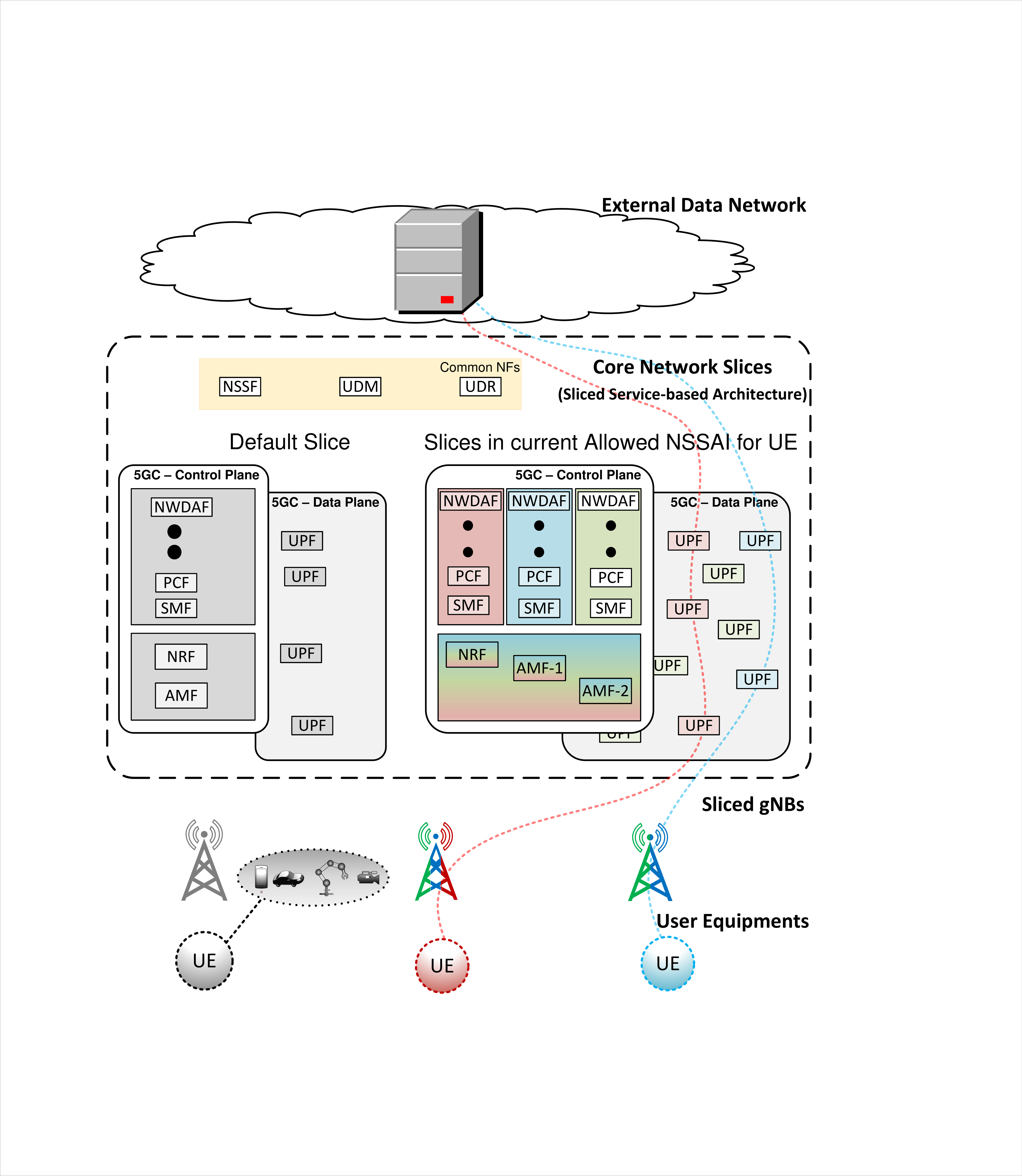}}
	\caption{Illustration of a sliced 5G network, showing inter-slice mobility 
		of a UE.}
	\label{NewArchitectureDiagram}
\end{figure}
The 3GPP defines a network slice as a logical network with specific network 
capabilities. A logical network or a slice consists of a set of (usually 
virtual) network functions (NFs). In SBA, an individual slice is 
identified through an S-NSSAI (Single-Network 
Slice Selection Identifier). A mobile network domain 
can have multiple slices configured, which are (collectively) termed as 
Configured NSSAI. A UE can subscribe to multiple slices in the network which 
are termed as Subscribed \linebreak S-NSSAIs. However, at any given time, 
the network allows 
a UE to connect to only a limited number of slices, which are identified as 
Allowed NSSAI. The network can usually serve the UE through a default slice, 
when 
the UE does not show its preference to any particular slice(s). A typical 
network-sliced SBA 
is shown in Fig. \ref{NewArchitectureDiagram}. Full definition 
of the NFs in SBA can be found in \cite{23501}.

\begin{figure}[tbp]
	\centerline{\includegraphics[scale=0.26]{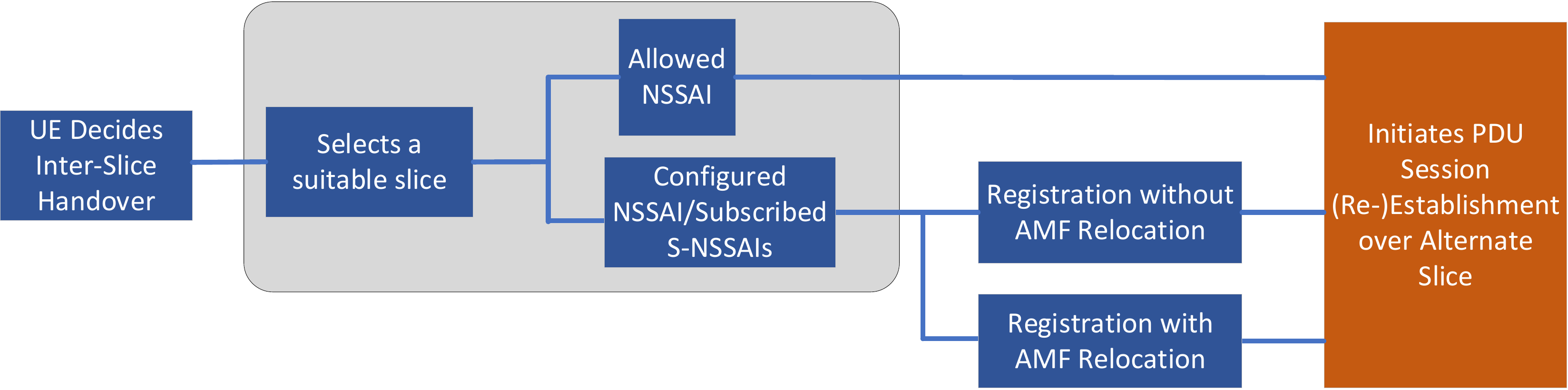}}
	\caption{Sequence of 3GPP standard operations when a UE decides to move 
	from one slice to another.}
	\label{BlockDiagram}
\end{figure}

When a UE wishes to (or is forced by the network to) change its slice, it 
has 
choice to select its alternate slice from the Allowed NSSAI. If the desired 
slice is not present in Allowed NSSAI, and it wishes to connect to a slice in 
the Configured NSSAI/Subscribed S-NSSAI, it is first required to carry out 
Registration to obtain its 
desired slices in Allowed NSSAI. 
The Registration process may in turn result in relocation 
of the current Access and Mobility Management Function (AMF) serving 
the UE (e.g., from AMF-1 to AMF-2 in Fig. \ref{NewArchitectureDiagram}).
Hence, the inter-slice mobility can occur in several forms, which we 
have discussed in our recent work in \cite{Sajjad2020}. In this paper, we 
consider a specific case when a UE has its 
desired alternate 
slice in Allowed NSSAI, either already present, or obtained after performing 
Registration. This process is shown in Fig. \ref{BlockDiagram}. 


%
According to the current specification, when a UE wishes to move from one 
slice to another, it is required to \linebreak(re-)establish its session over 
new slice 
for 
service continuation. The PDU Session (re-)establishment request for this 
purpose triggers the AMF to release the UE's session through Session Management 
Function (SMF) at previous slice 
\cite{23502}.  
Intuitively, this process is prone to significant service disruptions and 
packet drops. Hence, the inter-slice mobility of a UE requires smooth 
continuation 
of its ongoing session  among slices. 
\section{Proposed Solutions}
In this section, we present three approaches for inter-slice session 
continuity of IPv6 sessions. These solutions have several common operations and 
are hence 
collectively represented in Fig. \ref{solution1}. All these solutions 
consider 
the standard PDU Session establishment process \cite{23502} as a baseline, and 
enhance it 
either with MIPv6-RR/BU, inter-slice GTP tunnelling, or combination of both. 
\subsection{MIPv6-RR/BU-based Session Continuation}
\label{sec: solution1}
\begin{figure*}[htbp]
	\centerline{\includegraphics[scale=0.325]{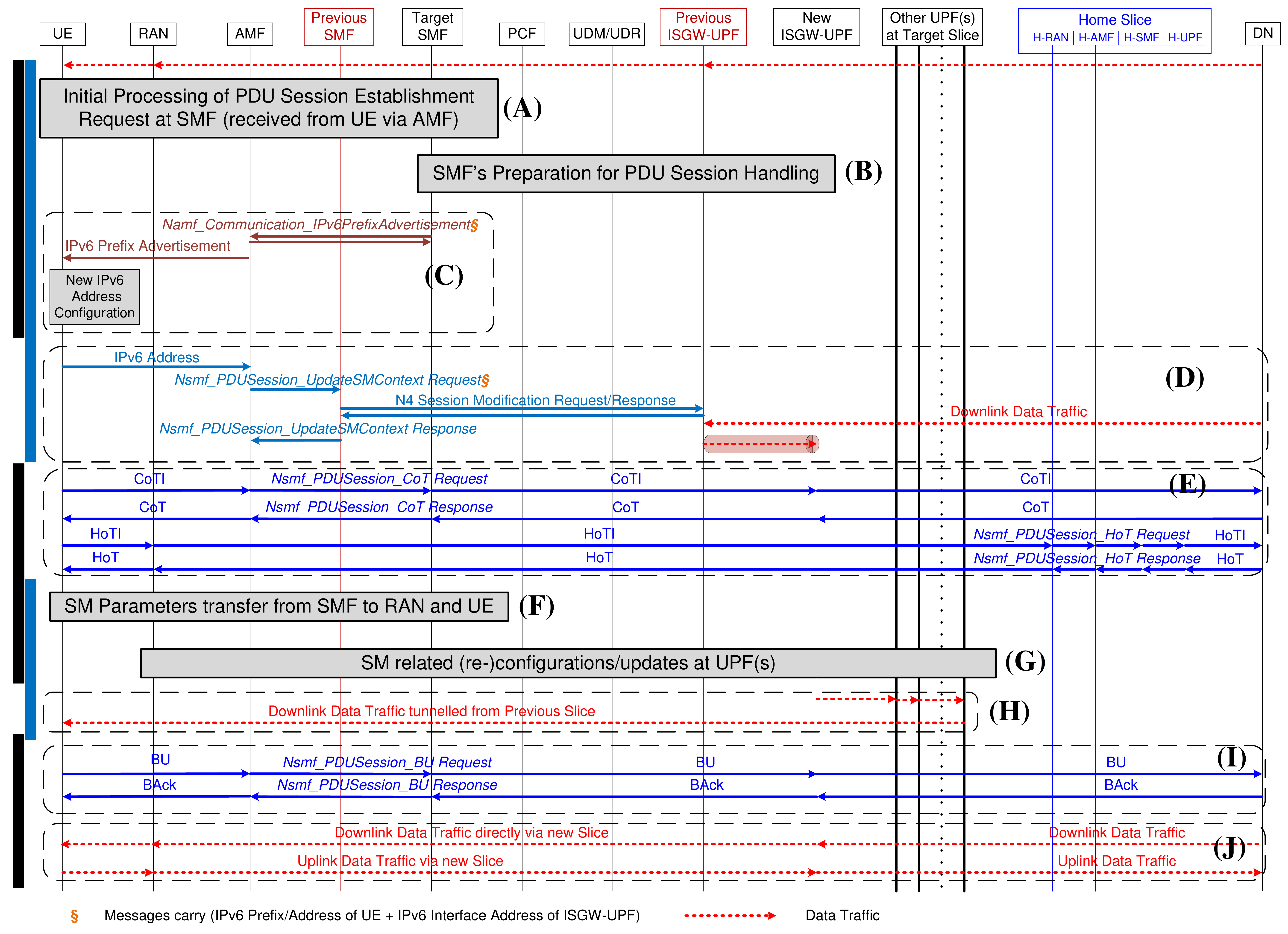}}
	\caption{Signalling sequence for proposed solutions. The signalling 
		sequence following the black scale on the left hand side represents 
		MIPv6-RR/BU solution (Steps \textbf{(A)-(C)}, \textbf{(E)-(G)}, 
		\textbf{(I)-(J)}). The 
		signalling sequence following the blue scale represents GTPv1-U 
		(Steps \textbf{(A)-(D)}, \textbf{(F)-(H)}), and the complete diagram 
		represents 
		the 
		hybrid 
		MIPv6/GTPv1-U 
		solution 
		(Steps \textbf{(A)-(J)}). For Steps \textbf{(A)}, \textbf{(B)}, 
		\textbf{(F)}, \textbf{(G)}, the detailed signalling 
		sequence is omitted, since it is based on the standard PDU Session 
		Establishment process \cite{23502}.} 
	\label{solution1}
\end{figure*}
Mobile IPv6 is developed to enable session continuity of a user when it moves 
out of its home network, and roams around different IP domains. The user 
remains reachable to its home network by updating it with its new (temporary) 
IPv6 address. The user thus continues to receive traffic at its new location 
via home network. However, this causes significant overheads \cite{makaya2008}, 
and hence the 
direct traffic exchange between user and the 
correspondent node (CN) from new network is enabled, as the user completes the 
Return Routability (RR) and Binding Update (BU) with the CN \cite{mipv6}.

The proposed MIPv6-RR/BU-based session continuation solution is based on the 
RR/BU procedures of MIPv6. Corresponding to home network in MIPv6, we propose 
a concept of \textit{“Home Slice”}, wherein the default slice of UE functions 
as its 
home slice. The home slice remains reachable to UE via its current RAN or AMF. 
The UE's subscription data remains available at its Home AMF (H-AMF) and 
Home SMF (H-SMF), even when the home slice is removed from the Allowed NSSAI. 
This 
allows any control plane signalling exchange between UE and DN over the home 
slice. The full protocol operation in MIPv6-RR/BU-based session continuation 
consists of Steps \textbf{(A)}-\textbf{(C)}, \textbf{(E)}-\textbf{(G)}, and 
\textbf{(I)}-\textbf{(J)}  
shown in Fig. 
\ref{solution1}, and are briefly described below.

\textbf{(A)} The UE, on deciding ISHO (and acquiring the desired slice 
in Allowed NSSAI), sends the standard “PDU 
Session Establishment 
Request” message to its current AMF. A new \textit{Request Type} ``Existing PDU 
Session Takeover Request" is proposed which indicates that the request message 
is about an 
ongoing session to be taken over from another slice. The AMF selects a suitable 
SMF for UE, and transfers the received request to 
SMF. 
The SMF, on checking its 
\textit{Request 
Type}, learns of an existing session from 
another slice. It then verifies the 
subscription information of UE, which it can retrieve from the Unified Data 
Management/Repository (UDM/UDR). After 
verification, it creates a Session Management (SM) context and responds 
to the AMF \cite{23502}.  

\textbf{(B)} The SMF starts preparation to 
handle (or takeover) the existing session. For this purpose, it performs three 
key standard operations: (a) It selects and establishes the policy association 
with a Policy 
Control Function (PCF) which provides policy rules.  (b) 
It selects the User Plane Functions (UPFs) which will handle the data 
plane traffic. The SMF also performs the N4 sessions establishment which 
enables SMF-UPF interaction with the 
selected UPF(s), and deploys policy rules on these UPF(s). (c) The 
SMF allocates the IP 
address for UE (i.e., an IPv6 prefix for an IPv6 PDU Session) \cite{23502}. 

\textbf{(C)} In 3GPP specifications, the allocated IP information (i.e., an 
IPv6 Prefix) 
is shared with UE after the standard 
``PDU 
Session Establishment" process completes \cite{23502}. For ISHO support, we 
propose that the 
SMF shares the IP information with UE as it is allocated. 
For this purpose, the SMF invokes the proposed 
\textit{IPv6PrefixAdvertisement} service operation of the standard 
\textit{Namf\_Communication} 
service with AMF. 
The AMF carries out the IPv6 Prefix 
Advertisement, and the UE can configure its new IPv6 
address using the received prefix through stateless IPv6 address 
autoconfiguration \cite{ipv6autoconfig}. 

\textbf{(E)} The UE having new IPv6 address is ready to carry out RR. It 
creates the 
standard MIPv6 messages -- Home 
Test Init (HoTI), and Care-of Test Init (CoTI). 
The UE sends the CoTI message to the DN via new 
slice, while the HoTI message is sent via Home Slice. The DN responds to both 
messages with CoT and HoT messages. 
New service operations, \textit{CoT} and \textit{HoT} are proposed to the 
standard \textit{Nsmf\_PDUSession} service, to enable transmission of CoTI/CoT 
and HoTI/HoT messages respectively between AMF and SMF (as shown in Fig. 
\ref{solution1}).


\textbf{(F)} The SMF, on the other hand, is also required to 
carry out the QoS and other resources set up at RAN and UE for proper 
handling of the PDU Session.
On successful configurations, the PDU 
Session Establishment Accept message is also delivered to the UE \cite{23502}. 

\textbf{(G)} The SMF carries out the N4 sessions modification with UPFs if QoS 
configurations and resources set up at RAN and UE require 
parameters re-adjustments at the core network \cite{23502}. 

\textbf{(I)} After Step \textbf{(F)}, provided that RR
completes successfully, the UE can send the MIPv6 standard Binding Update (BU)
message to the DN 
over the new slice. The DN, on receiving and verifying  
the BU, responds the UE with Binding Acknowledgement (BAck) message. 

\textbf{(J)} The acceptance of BU in Step \textbf{(I)} also triggers the DN 
to direct 
the ongoing traffic 
of the UE over its new slice. 
\subsection{GTPv1-U based Session Continuation}
\label{sec: solution2}
The GTPv1-U is a default user plane 
tunnelling protocol in 3GPP specifications for 5G \cite{23501}. It
handles tunnelling between two UPFs, and also between a UPF and a 5G access 
network node (e.g., a gNB). In GTPv1-U, a 
datagram (e.g., an IP datagram) is encapsulated in a GTPv1-U header and then in 
a 
UDP/IP header \cite{gtpv1u}. 

In order to support GTP tunnelling among slices, the proposed \textit{ISGW} 
operates on a 
UPF (hence termed as ISGW-UPF).
The ISGW-UPF establishes the GTPv1-U tunnel with its peer ISGW-UPF in the 
Target Slice. This solution consists of Steps \textbf{(A)}-\textbf{(D)} and 
\textbf{(F)}-\textbf{(H)} 
as shown 
in 
Fig. \ref{solution1}. In addition to the details described in Section 
\ref{sec: solution1}, following additional operations take place in each step. 

\textbf{(A)} When the AMF receives the PDU Session Establishment Request, 
it is required to coordinate with both slices (i.e., previous slice and the new 
slice). It coordinates with the new slice first in Step \textbf{(A)}, and if 
the new slice is willing to cater the UE's session, then the AMF (later on in 
Step \textbf{(D)}) coordinates 
with previous slice to request it to divert the UE's traffic towards new slice. 

\textbf{(B)} The N4 sessions establishment in Step \textbf{(B)} includes 
configurations 
for 
ISGW-UPF as well. This prepares the ISGW-UPF to cater the UE's traffic in both 
uplink/downlink directions to/from 
another slice. 

\textbf{(C)} The SMF also 
includes the IPv6 address of the ISGW-UPF interface which will receive 
the incoming traffic from the previous slice, when it invokes the proposed
\textit{Namf\_Communication\_IPv6PrefixAdvertisement} service 
operation with AMF (Fig. \ref{solution1}). 

\textbf{(D)} The UE, after configuring its new IPv6 address in Step 
\textbf{(C)}, 
communicates 
it to AMF. The AMF provides the IPv6 Address to previous slice through 
\textit{Nsmf\_PDUSession\_UpdateSMContext} Request. The IPv6 
interface address of the ISGW-UPF of
new slice is also included. The SMF performs N4 sessions modification with 
ISGW-UPF to enable GTPv1-U tunnel establishment with ISGW-UPF in the target 
slice. The UE's incoming traffic is subsequently
tunnelled to the ISGW-UPF in the target 
slice. 
%




\textbf{(H)} The new slice starts receiving traffic from the previous 
slice, and delivers it to the UE, as shown in Fig. \ref{solution1}. 
%
%
%
%
%
%
\subsection{MIPv6/GTPv1-U based Session Continuation}
The MIPv6/GTPv1-U solution is the combination of MIPv6-RR/BU and the 
GTPv1-U 
based session continuation solutions. In MIPv6/GTPv1-U,
the incoming traffic of UE at previous slice is tunnelled to new slice 
temporarily. 
As the AMF detects BAck (in Step \textbf{(H)}), it can 
invoke the standard \textit{Nsmf\_PDUSession\_ReleaseSMContext} 
service operation with SMF at previous slice \cite{23502} to release the 
tunnel. The full operation 
of MIPv6/GTPv1-U consists of
Steps \textbf{(A)}-\textbf{(J)}, as 
described 
in Section \ref{sec: solution1} and 
the Section \ref{sec: solution2}, and shown in Fig. \ref{solution1}. 
\section{Analytical Modelling}
In this section, through analytical modelling, we compare the 
performance of proposed solutions with the baseline PDU Session 
Establishment process \cite{23502}. Several notations used in analytical 
modelling are defined in Table \ref{tab1}. 
\subsubsection*{System Modelling}
\begin{figure}[htbp]
	\centerline{\includegraphics[scale=0.125]{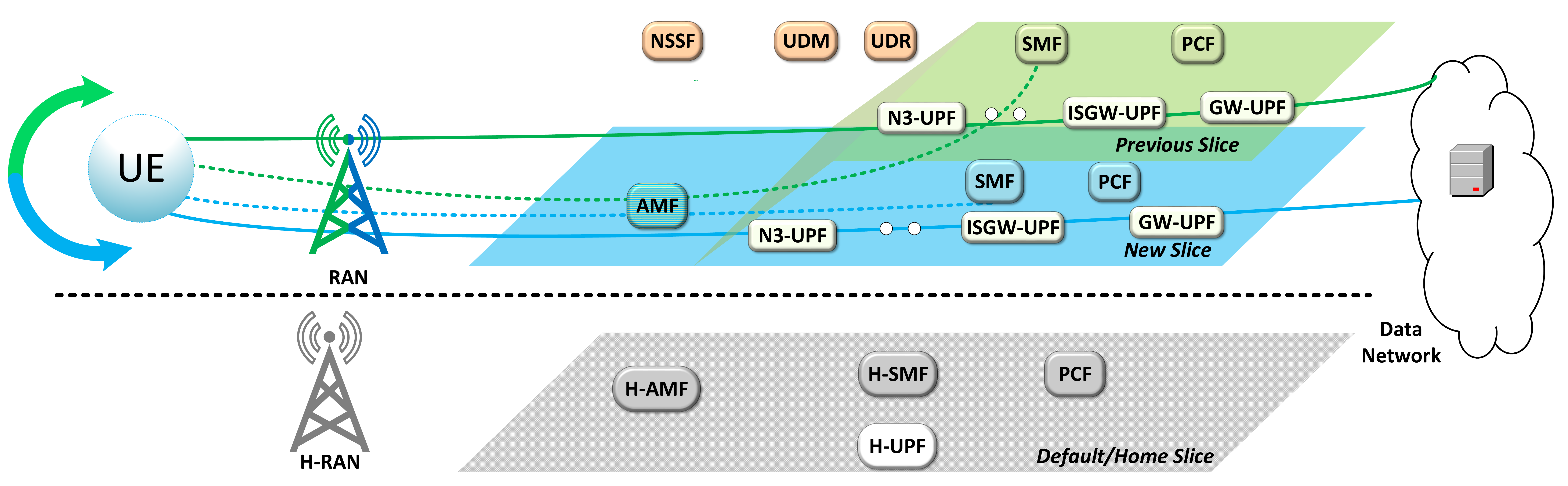}}
	\caption{System Model}
	\label{SystemModel}
\end{figure}

For comparative analysis, we consider the system model 
shown in Fig. \ref{SystemModel}. 
The core network NFs are assumed to be virtual NFs deployed at commodity 
hardware, which can be dedicated or shared among virtual 
NFs. 
%
The data plane consists of UPFs, which include a GW-UPF, the proposed 
ISGW-UPF, an N3-UPF which terminates the N3 reference point with RAN 
\cite{23501} and other 
UPFs as required. A common data plane will normally configure 2 or 3 
UPFs \cite{upfdraft2019}. 
At RAN, a sliced gNB is considered which 
supports both current and target slice of UE. Since the signalling related to 
the PDU Session management 
is handled from the 5GC, the primary focus of our work remains on the 
inter-action between UE, the 5GC and the DN.
%
%
\subsubsection*{Mobility Modelling}
The inter-slice mobility does not necessarily require 
physical mobility of the UE. We assume that the current gNB supports both 
slices, and the UE does not change the gNB during the inter-slice handover. It 
is worth remarking that should the inter-slice handover occurs as a result of 
UE's movement towards another gNB or different access technology, then the 
horizontal and 
vertical handover management procedures will also be required, respectively. 
These will 
precede the 
inter-slice handover operation.
Both these cases represent complex mobility 
management scenarios, and 
require further investigations \cite{Sajjad2020}. 
%
%
\subsubsection{ISHO Delay and ISHO Interval}
The ISHO delay is defined as the time interval from the 
instance 
when the UE receives the last downlink packet at previous slice, to the 
instance 
when it receives the first downlink packet (or can send the first uplink 
packet) from new slice. 
%
%
We also define the total ISHO interval, which is the duration from when the UE 
starts the ISHO process until it completes. The ISHO delay and the total ISHO 
Interval are represented respectively 
as $T^{(\cdot)}$, and $L^{(\cdot)}$. The notation $(\cdot)$, hereinafter, is 
used to represent either of the proposed schemes (i.e., MIPv6-RR/BU as $Mipv6$, 
GTPv1-U as $Gtp$, 
MIPv6/GTPv1-U as $Mipv6-Gtp$) as well as the baseline PDU Session Establishment 
procedure (as $3gpp$). The expressions for $T^{(\cdot)}$ and $L^{(\cdot)}$ are 
given in Table \ref{tab2}, where the terms 
$T_A$ to $T_J$ represent the delays 
associated to Steps \textbf{(A)} to \textbf{(J)} in Fig. \ref{solution1}. These 
delays are sum 
of the 
transmission delays ($T_{x,y}$) between a node (or an NF) $x$ and another node 
(or NF) $y$, and processing delays ($PD_x$) at any node 
(or NF) $x$ \cite{makaya2008}. These are simple to formulate, for instance, 
$T_C$ 
can be given as, 
\begin{equation}
\begin{split}
T_{C}=T_{SMF, AMF} + PD_{AMF} + T_{AMF, UE} + PD_{UE}
\end{split}
\end{equation}

Similar expressions can also be formulated for $T_{Sec-Auth}$ and 
$T_{STD-IPv6Config}$ which represent the standard ``Secondary 
Authentication/Authorization" for UE by DN and  ``IPv6 Advertisement" by 
SMF to UE via UPF(s) \cite{23502} respectively.  
\begin{table}[ht!]
	\caption{ISHO delay and ISHO Interval Expressions}
	\begin{center}
		\begin{tabular}{p{8.5cm}}
			\hline
			$T^{3gpp}=T_{A} + T_{B} + T_{Sec-Auth}+T_{F}+T_G+ 
			T_{STD-IPv6Config} +T_{J}$  
			\\ 
			$L^{3gpp}=T^{3gpp}$\\
			$T^{Mipv6}=max\{T_{E}, T_{D}\}+T_{I}+T_{l}+T_{H}$\\
			$L^{Mipv6}=T_{A}+T_{B}+T_{C} +T^{Mipv6}$\\ 
			$T^{Gtp}=max\{T_{D},T_{F}+T_{G}\}+T_{J}$\\
			$L^{Gtp}=T_{A}+T_{B}+T_{C}+T^{Gtp}$\\
			$T^{Mipv6-Gtp}=min\Big\{max\{T_{D}, T_{F}+T_{G}\}+T_{J}, 
			max\{T_{E}, T_{F}\} + T_{I} + T_{H}\Big\}$\\
			$L^{Mipv6-Gtp}=T_{A}+T_{B}+T_{C}+max\Big\{max\{T_{D}, T_{F}+T_{G}\}
			+T_{J}, max\{T_{E}, T_{F}\} + T_{I} + T_{H}\Big\}$\\
			\hline
		\end{tabular}
		\label{tab2}
	\end{center}
\end{table}
\subsubsection{Core Network Resources Overhead}
The core network resource overhead represents the overall resource consumption 
by a flow at the data plane of a slice. These resources include the fraction of 
CPU resources ($c$), and the fraction of link bandwidth resources ($b$), 
allocated to an ongoing flow \cite{Ye2019}. Their respective overheads are 
represented as $CR^{(\cdot)}$ and $BR^{(\cdot)}$. Assuming that these resources 
are statically allocated, $CR^{(\cdot)}$ and $BR^{(\cdot)}$ can be given as, 
\begin{equation}
\begin{split}
CR^{(\cdot)}= \omega_p \cdot c_p \cdot N_{UPF}^{p*} + 
\omega_p \cdot c_n \cdot N_{UPF}^{np*} 
+ \omega_n \cdot c_n \cdot N_{UPF}^{n*}
\end{split}
\end{equation}
\begin{equation}
\begin{split}
BR^{(\cdot)}= \omega_p \cdot c_p \cdot (N_{UPF}^{p*} -1)+ 
\omega_p \cdot c_n \cdot (N_{UPF}^{np*} - 1) \\+ \omega_n
\cdot c_n \cdot (N_{UPF}^{n*} - 1)
\end{split}
\end{equation}

\begin{table*}[h!]
	\caption{System Parameters}
	\begin{center}
		\begin{tabular}{p{1.8cm}p{10.9cm}p{3.8cm}}
			\hline
			\textbf{Notation}& \textbf{Description} & \textbf{Default Values} 
			\\ \hline 
			$T_{nf, nf}$ & Avg. Transmission delay between two 
			control plane SBA NFs 
			& 1 ms\\
			$T_{x, y}$& Avg. Transmission delay between a non-SBA node and 
			another node (or an SBA NF)  & 	Manifold, defined in Section 
			\ref{sec: ISHOIntervalDelayAnalysis}	\\
			$PD_{nf}$& Avg. Processing delay at SBA CP NF &  1 ms\\
			$PD_{x}$ & Avg. Processing delay at non-SBA node $x$ & Manifold, 
			defined in Section \ref{sec: ISHOIntervalDelayAnalysis}  \\
			$N_{UPF}^{p}, N_{UPF}^{n}$ & Maximum No. of UPFs deployed at 
			previous/new slice &  3\\
			$N_{UPF}^{p*}, N_{UPF}^{n*}$ & No. of UPFs currently handling UE 
			traffic at previous/new slice &  1, $N_{UPF}^{n}$\\
			$N_{UPF}^{np*}$ & No. of UPFs at new Slice handling incoming 
			traffic from previous &  $N_{UPF}^{n}$\\
			$N_{UPF}^{p'}, N_{UPF}^{n'}$ & Total No. of UPFs between GW-UPF and 
			ISGW-UPF at 
			Previous, new Slice &  1, 0\\
			$c_p$, $c_n$& Fraction of CPU processing resources allocated to a 
			flow at $N_{UPF}^{p}, N_{UPF}^{n}$ & 100 packets/sec.\\
			$PC_{nf}, PC_x$& Avg. Processing Cost of a control packet at an NF 
			in 
			SBA, or non-SBA node $x$ & 1, 5 
			\\
			$TC_{nf,nf}$& Avg. Transmission cost between two NFs at SBA &  1\\
			$TC_{x, y}$& Avg. Transmission cost between a node (or NF) $x$ 
			to another node (or NF) $y$ & 2 \\
			%
			%
			%
			%
			%
			%
			%
			%
			\hline
		\end{tabular}
		\label{tab1}
	\end{center}
\end{table*}
$\omega_p$ and $\omega_n$ are the ratios of traffic from 
indirect path (i.e., via previous slice) and the direct path (i.e.,
via new slice) respectively. 

For $CR^{3gpp}$ and $BR^{3gpp}: \omega_p = 0; \omega_n=1$, $N_{UPF}^{p*} 
= 0$,  
$N_{UPF}^{np*}= 0$,  $N_{UPF}^{n*} = N_{UPF}^{N}$.
\smallskip

For $CR^{Mipv6}$ and $BR^{Mipv6}: \omega_p = 0; \omega_n=1$, 
$N_{UPF}^{p*} = 
0$,  $N_{UPF}^{np*}= 0$,  $N_{UPF}^{n*} = N_{UPF}^{N}$.
\smallskip

For $CR^{Gtp}$ and $BR^{Gtp}: \omega_p = \omega_n = 1$, 
$N_{UPF}^{p*} = 
N_{UPF}^{p'}$,  
$N_{UPF}^{np*}= [N_{UPF}^{N} - N_{UPF}^{n'}] $,  $N_{UPF}^{n*} = 0$.
\smallskip

For $CR^{Mipv6-Gtp}$ and $BR^{Mipv6-Gtp}: \omega_p = 1 - 
\omega_n$, 
$N_{UPF}^{p*} = N_{UPF}^{p'}$,  $N_{UPF}^{np*}= [N_{UPF}^{N} - N_{UPF}^{n'}]$,  
$N_{UPF}^{n*} = N_{UPF}^{N}$.
\subsubsection{Signalling Cost}
Signalling cost consists of transmission cost ($TC_{x, y}$) between any node 
(or NF) $x$ and another node (or NF) $y$ and processing cost ($PC_x$) at any 
node (or NF) $x$ \cite{makaya2008}. From Fig. \ref{solution1}, these can be 
formulated for each solution, for example, for Step \textbf{(C)} as,  
\begin{equation}
\begin{split}
SC_{C}=2 \cdot TC_{NF, NF} + 2 \cdot PC_{NF} + TC_{NF, UE} + PC_{UE}
\end{split}
\end{equation}
\section{Results}
For comparison, the default values of system parameters are shown in Table 
\ref{tab1}. Most of these are based on \cite{CostAnalysisLee2010, moneeb2016, 
makaya2008, JainPIMRC2018, LteFrameworkPrados2020}.
\begin{figure*}[htb!]
	\begin{multicols}{4}
		\includegraphics[width=\linewidth]{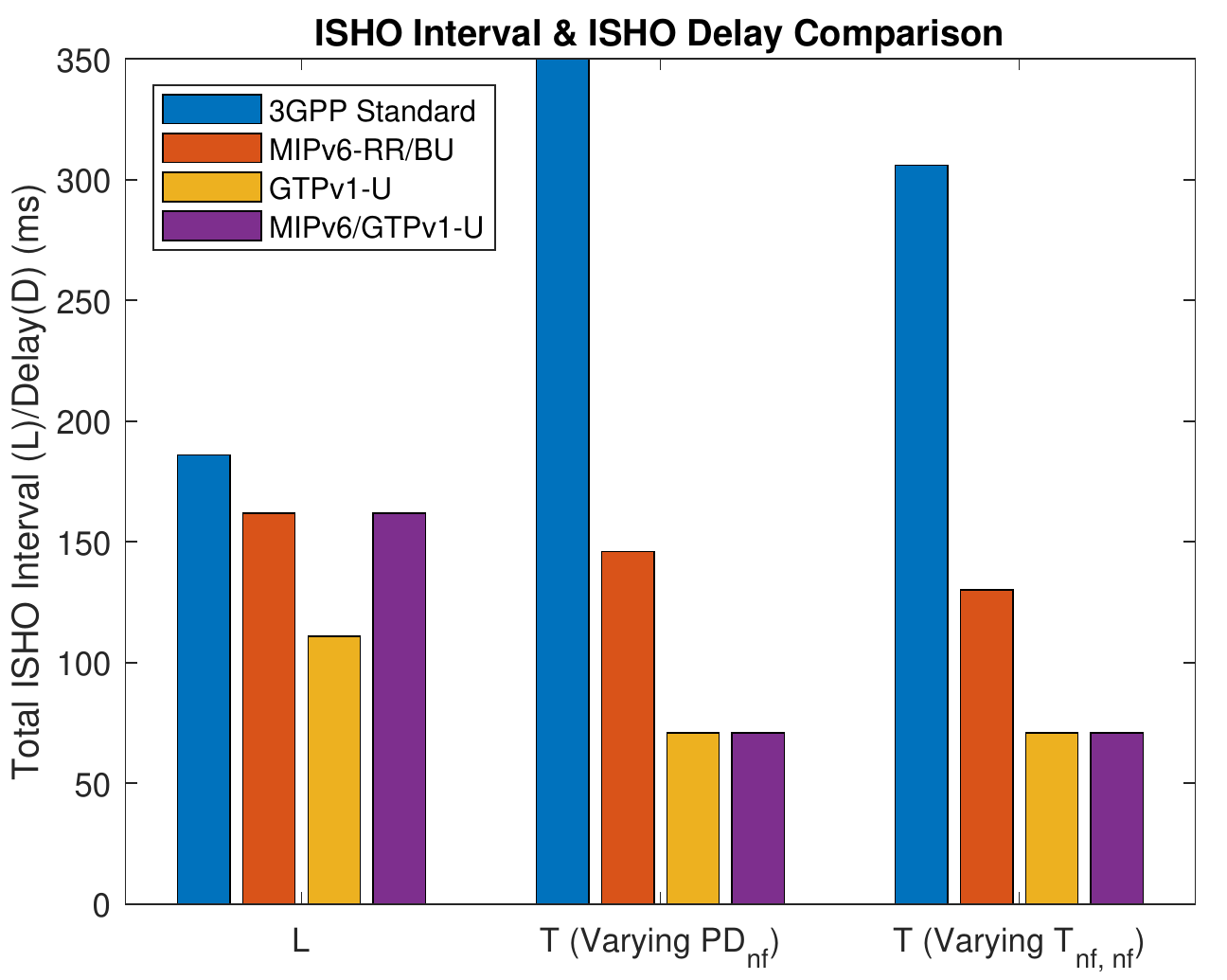}
		\caption{ISHO Interval and ISHO Delay comparisons.}
		\label{fig: ISHOIntervalAndDelayComparison}
		\includegraphics[width=\linewidth]{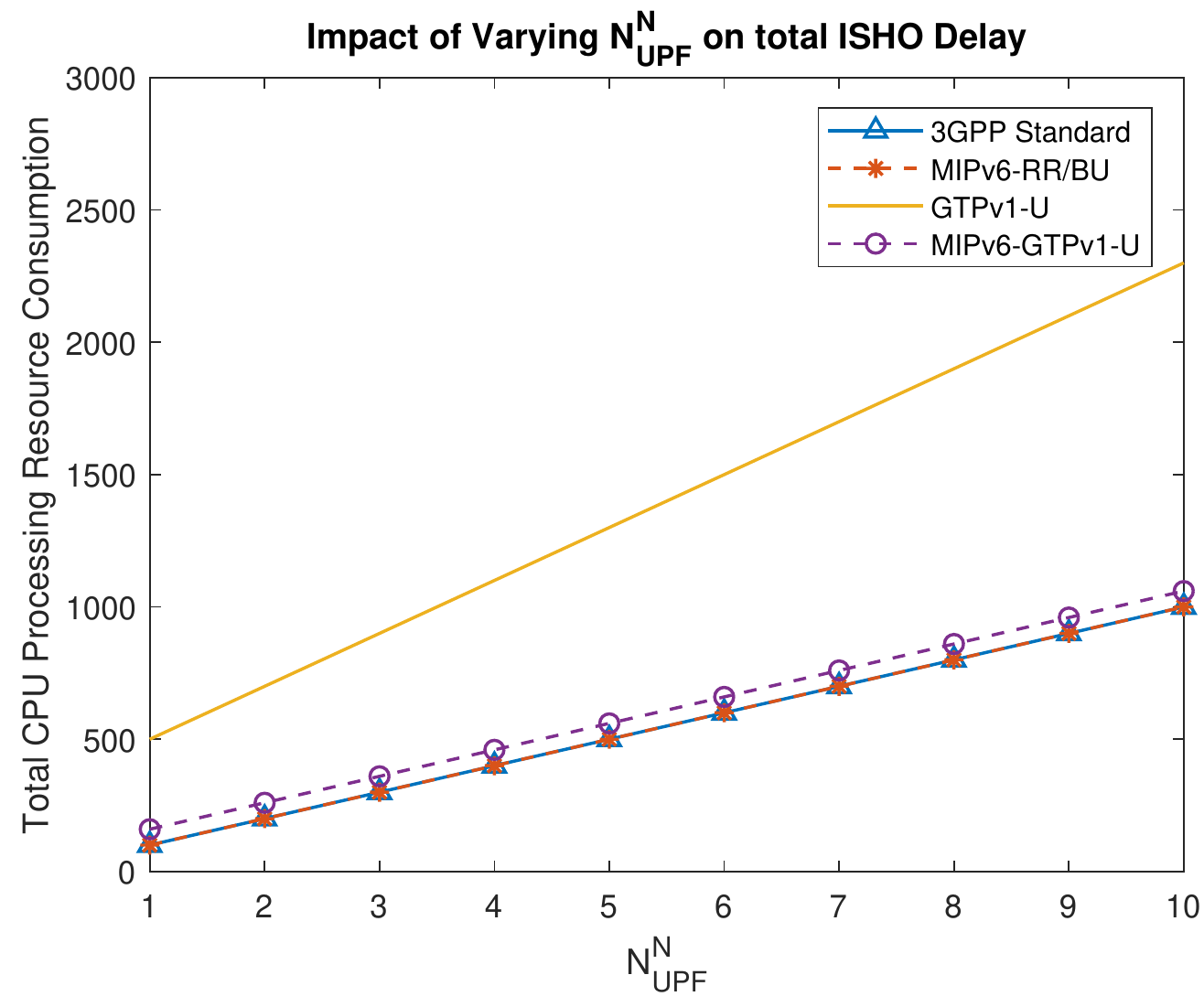}
		\caption{Core network resources overhead at data plane (Varying 
			$N_{UPF}^N$).}
		\label{fig: ResourceOverheadVaryingNNupf}
		\includegraphics[width=\linewidth]{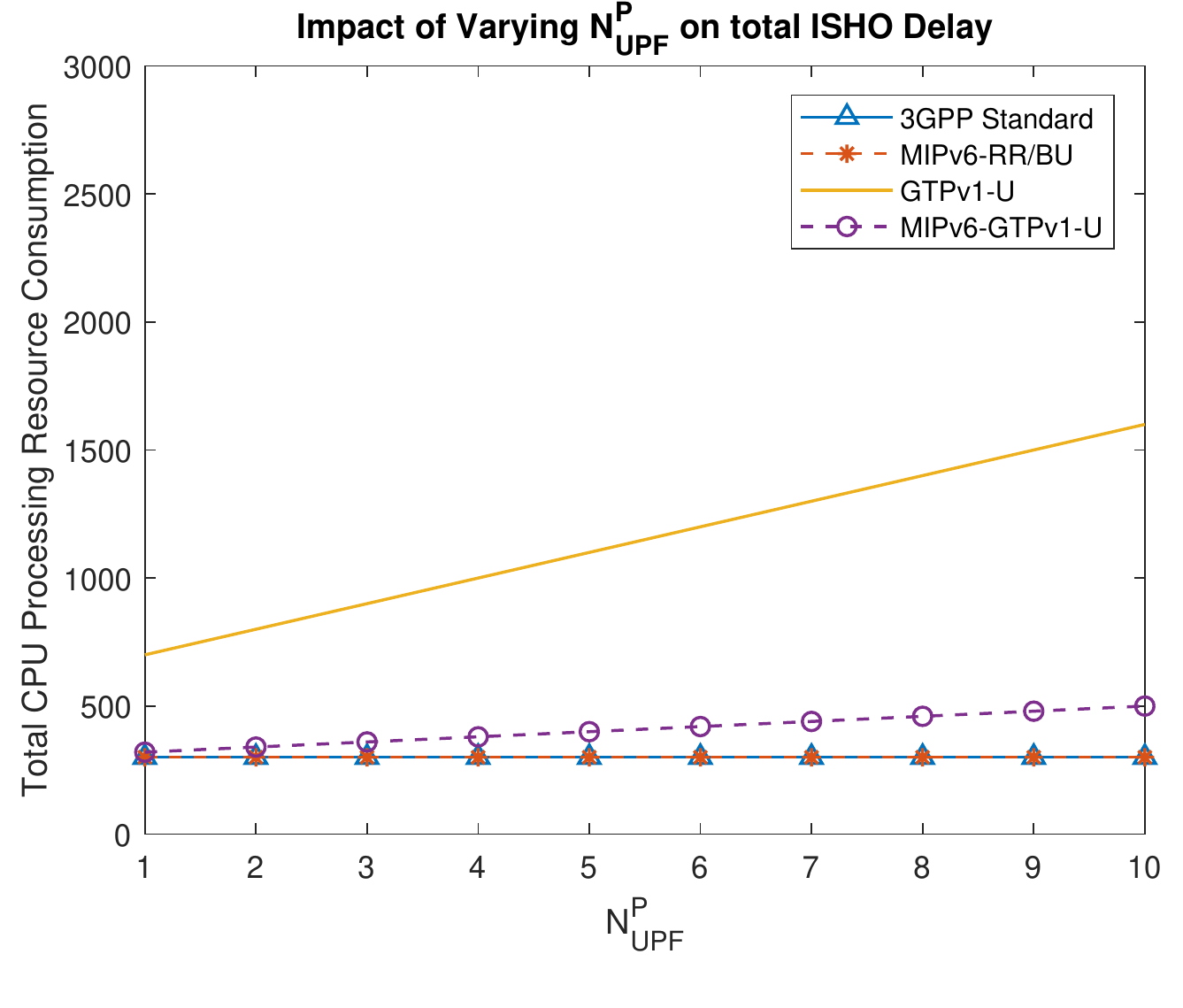}
		\caption{Core network resources overhead at data plane (Varying  
			$N_{UPF}^P$).}
		\label{fig: ResourceOverheadVaryingNPupf}
		\includegraphics[width=\linewidth]{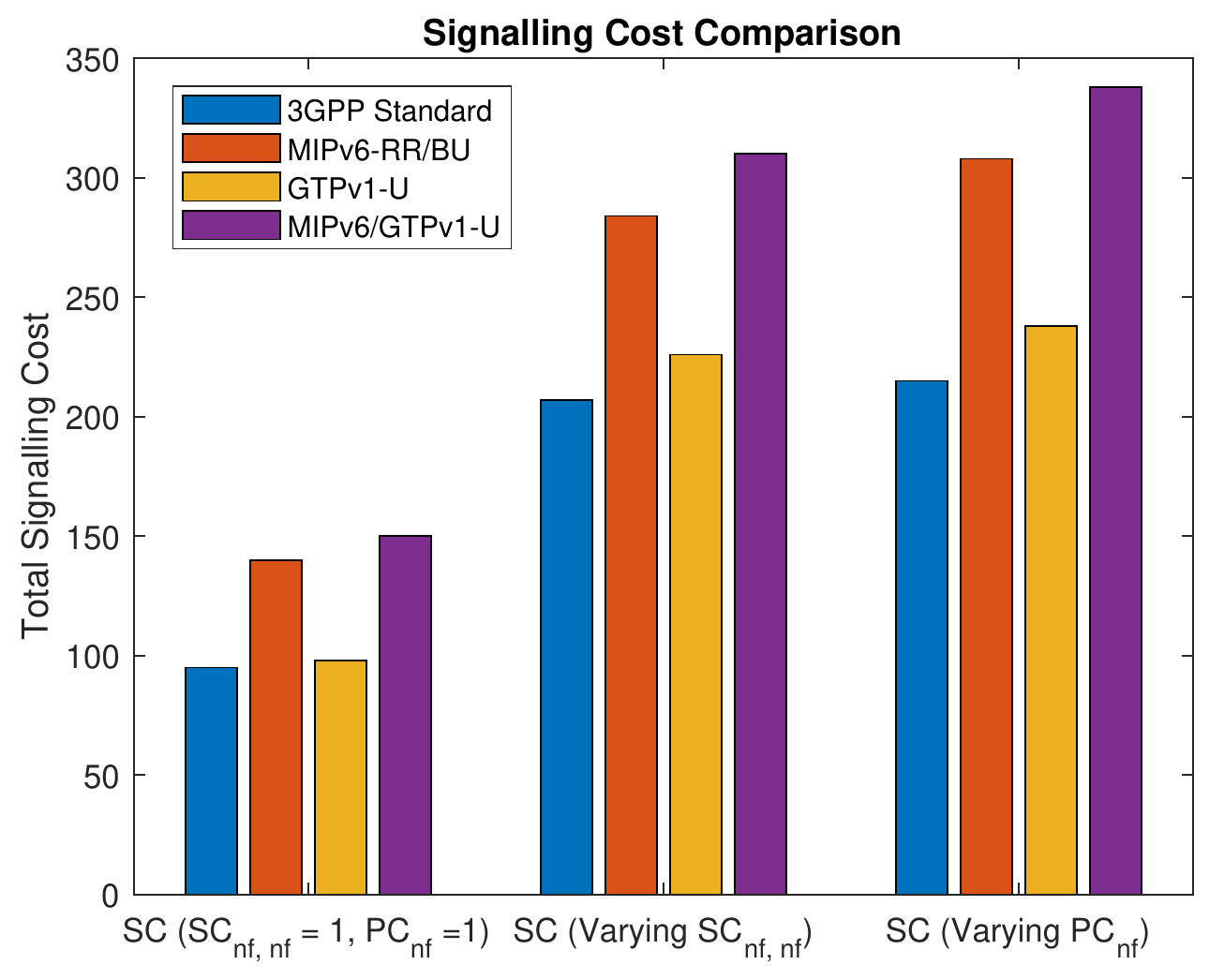}
		\caption{Signalling cost comparison (Varying $PC_{nf}$ and $SC_{nf, 
		nf}$).}
		\label{fig: SignallingCosts}
	\end{multicols}
\end{figure*} 
\subsubsection{ISHO Interval and ISHO Delay Analysis}
\label{sec: ISHOIntervalDelayAnalysis}
We study the impact of varying 
characteristics of the SBA slices in terms of the $T_{nf, nf}$ and $PD_{nf}$. 
The lower and higher values of these parameters characterize the control plane 
virtual NFs deployments at shared and dedicated hardware with shared and 
dedicated resources respectively.  
Other parameter values are set as, $T_{ue, amf}=5 ms$, $T_{ran, hamf}= 3 ms$, 
$T_{gw, dn} = 5 ms$, $T_{upf, upf}= 2 ms$, $T_{smf, upf}=2 ms$. The $PD_{x}$ 
for any non-SBA node $x$, and $PD_{upf}$ are set to $2 ms$. These non-SBA nodes 
include 
UE, DN, and gNB.  
As shown in Fig. \ref{fig: ISHOIntervalAndDelayComparison}, for total ISHO 
interval, 
with default $T_{nf, nf}$ and $PD_{nf}$ values, only 
the GTPv1-U 
solution shows notable reduction of about 40\% compared to the standard 3GPP 
process. However, all the proposed solutions achieve significant reduction for 
ISHO delay. For $PD_{nf}$ values varied from 1 to 5 $ms$ ($T_{nf, nf}=1 ms$), 
the MIPv6-RR/BU can achieve approximately 60\% reduction 
in ISHO delay compared to the standard 3GPP process. On the other hand, the 
GTPv1-U based solutions (i.e., GTPv1-U and MIPv6/GTPv1-U) achieve up to 80\% 
reduction in the overall delay. For 
$T_{nf, nf}$ values varied from 1 to 5 $ms$ ($P_{nf}=1 ms$), the MIPv6-RR/BU 
and 
GTPv1-U (and MIPv6/GTPv1-U) achieve 58\% and 76\% reduction in ISHO delay 
respectively, compared to the standard 3GPP process.
%
%
\subsubsection{Resource Overhead Analysis}
For resource overhead analysis, we consider a worst case scenario, where an 
ISGW-UPF is deployed at N3-UPF at previous slice, and at GW-UPF at the new 
slice. Hence, $N_{UPF}^{p*}=N_{UPF}^P$, and $N_{UPF}^{n*} = 
N_{UPF}^{np*}=N_{UPF}^N$. For MIPv6/GTPv1-U, we set $\omega_p=0.2$ 
\cite{CostAnalysisLee2010}. 
Multiple UPFs, especially for geographically 
widespread slices (e.g., for rate limiting, executing different QoS policies in 
different segments 
of the core network) are also likely. Accordingly, we vary 
$N_{UPF}^P$ and $N_{UPF}^N$, to analyze
the 
overall resource overhead. In this analysis, we only consider the 
$CR^{(\cdot)}$, since the $BR^{(\cdot)}$ for each scheme
is of the same proportion. Fig. \ref{fig: ResourceOverheadVaryingNNupf} shows 
that, with varying $N_{UPF}^N$,  the MIPv6-RR/BU incurs 
minimal resource overheads compared to GTPv1-U and MIPv6/GTPv1-U solutions. In 
fact, the resource overhead of MIPv6-RR/BU is similar to the standard 3GPP 
solution, since neither of these solutions require resources from 
previous slice due to direct communication establishment between UE and DN. 
The MIPv6/GTPv1-U solution, which only requires 
previous slice resources temporarily, shows significant reduction in resource 
consumptions, compared to the GTPv1-U solution. On the other hand, varying 
$N_{UPF}^P$, as shown in Fig. \ref{fig: ResourceOverheadVaryingNPupf},  the 
MIPv6/GTPv1-U for lower 
$N_{UPF}^P$ incurs approximately similar resource overheads as 3GPP standard 
process
and 
MIPv6-RR/BU solutions respectively. 
%
%
%
\subsubsection{Signalling Cost Analysis}
The signalling cost metric aptly represents the control plane traffic overhead 
incurred by each scheme. 
For the signalling cost comparison, it 
is clear from Fig. \ref{fig: SignallingCosts} that all the proposed solutions 
achieve session 
continuation at the cost of additional signalling. Among these, the GTPv1-U 
based session continuation incurs only 3\% additional costs compared to the 
standard 3GPP process, and up to 10\% additional costs for higher $TC_{nf, 
nf}$ or $PC_{nf}$ values. 
The MIPv6-RR/BU solution incurs about 35-45\% costs, while
MIPv6/GTPv1-U incurs between 55-60\% additional costs. 

\section{Conclusion}
In this paper, we have proposed and analyzed three solutions to achieve session 
continuation among slices in 3GPP SBA. Among the proposed solutions, the 
GTPv1-U based solution incurs minimal latencies, and lower signalling costs. 
However, it incurs significantly higher resource 
overheads. The MIPv6-RR/BU, on the 
other hand, promises comparatively lower resource overhead 
costs, however, incurs higher latencies and signalling costs. Their combination 
MIPv6/GTPv1-U achieves lower latencies, and also improves resource overhead 
costs. However, it incurs significantly high signalling costs. 
This demonstrates the need for further 
investigations on the problem of ISHOs, which, given high reliance of conceived 
5G use cases on network slicing, is a significant problem. 

\vspace{12pt}

\end{document}